\documentclass{article}

\usepackage{PRIMEarxiv}
\usepackage[utf8]{inputenc} %
\usepackage[T1]{fontenc}    %
\usepackage{hyperref}       %
\usepackage{url}            %
\usepackage{booktabs}       %
\usepackage{amsfonts}       %
\usepackage{nicefrac}       %
\usepackage{microtype}      %
\usepackage{lipsum}
\usepackage{fancyhdr}       %
\usepackage{graphicx}       %
\usepackage{subcaption}
\usepackage{caption}       %
\graphicspath{{figures/}}     %
\usepackage[table,xcdraw]{xcolor}
\usepackage{array}
\usepackage{colortbl}
\usepackage{placeins}
\usepackage{soul,color}
\usepackage{amsmath}
\usepackage{multirow} 
\usepackage{pifont}
\title{TCSP 2.0: Template Based Crystal Structure Prediction with Improved Oxidation State Prediction and Chemistry Heuristics
\thanks{\textit{\underline{Citation}}: 
\textbf{L.Wei et al. TCSP 2.0: Templated based CSP... 15 Pages.... DOI:000000/11111.}} 
}

\author{
}

\author{%
  Lai Wei$^1$, Rongzhi Dong$^1$, Nihang Fu$^1$, Sadman Sadeed Omee$^1$,  Jianjun Hu$^{1}$\thanks{Corresponding author: jianjunh@cse.sc.edu}\\
  $^1$Department of Computer Science and Engineering, University of South Carolina, Columbia, SC, USA\\
  \texttt{jianjunh@cse.sc.edu}
  }

\begin{document}
\maketitle

\begin{abstract}
Crystal structure prediction remains a major challenge in materials science, directly impacting the discovery and development of next-generation materials. We introduce TCSP 2.0, a substantial evolution of our template-based crystal structure prediction framework that advances predictive capabilities through synergistic integration of several key techniques into its major components.
Building upon TCSP 1.0's template-matching foundation, this enhanced version implements three critical innovations: (1) replacement of pymatgen with deep learning-based BERTOS model for oxidation state prediction with superior performance, (2) implementation of sophisticated element embedding distance metrics for improved chemical similarity assessment, and (3) development of a robust majority voting mechanism for space group selection that reduces prediction uncertainty.
TCSP 2.0 also expands its template base by incorporating template structures from Materials Cloud, C2DB, and GNoME databases alongside the original Materials Project repository, creating a more comprehensive structural foundation. Rigorous validation across 180 diverse test cases of the CSPBenchmark demonstrates TCSP 2.0's exceptional performance, achieving 83.89\% space-group success rate and 78.33\% structural similarity accuracy for top-5 predictions, substantially outperforming both its predecessor and the competing modern CSP algorithms including CSPML and EquiCSP.
The related open-source code can be accessed freely at \url{https://github.com/usccolumbia/TCSP}.

\end{abstract}


\section{Introduction}
Crystal structure prediction (CSP) represents a cornerstone challenge in materials science and chemistry, which plays a crucial role in the discovery and development of novel materials. Accurate prediction of crystal structures is fundamental to understanding the physical, chemical, and electronic properties of a material, enabling the design of materials with customized functionalities for applications in energy storage, catalysis, and electronics \cite{oganov2018crystal,yokoyama2024crystal,leitherer2023automatic}. Over the years, significant progress has been made in CSP through a variety of approaches, ranging from empirical to machine learning-based algorithms \cite{graser2018machine,li2024machine,zhu2022materials}. 
A wide range of CSP algorithms have been introduced to tackle this challenge, including random search \cite{pickard2006high,pickard2011ab}, simulated annealing \cite{kirkpatrick1983optimization}, basin hopping \cite{wales1997global}, minima hopping \cite{goedecker2004minima}, evolutionary algorithms (EA)\cite{oganov2006crystal,lyakhov2013new}, particle swarm optimization (PSO) \cite{wang2010crystal,zhang2017computer}, Bayesian optimization (BO) \cite{yamashita2018crystal}, and look-ahead based on quadratic approximation (LAQA) \cite{terayama2018fine}. Despite their success, these methods face limitations when it comes to scalability and the exploration of high-dimensional compositional and structural spaces. Among these, software tools such as USPEX \cite{oganov2006crystal} and CALYPSO \cite{wang2012calypso}, have implemented various genetic manipulations and provided valuable insights into crystal structure prediction, but have often been limited by their limited scalability due to the computational expense of the Density Funational Theory (DFT) calculations and the need for human expertise in the setup of the problem. 

To address the computational bottlenecks of traditional CSP methods, machine learning-based approaches have gained traction in recent years. For example, CrySPY \cite{yamashita2021cryspy} introduced a Gaussian process regressor \cite{seeger2004gaussian} as a machine learning energy calculator, enhancing computational efficiency by prioritizing promising candidates through Bayesian optimization \cite{mockus1991bayesian}. GN-OA algorithms \cite{cheng2022crystal} utilize a graph neural network (GN) to model correlations between crystal structures and formation energy, coupled with an optimization algorithm (OA) to expedite the search for structures with the lowest formation enthalpy. These methods have been evaluated with graph neural network potentials such as MEGNet \cite{treacher2021megnet}, which has been combined with random search (RAS), Bayesian optimization (BO), and Particle Swarm Optimization (PSO). Building on this, ParetoCSP \cite{omee2024crystal} advances the GN-OA framework by incorporating a multi-objective genetic algorithm (MOGA) \cite{murata1995moga} with stronger search capability and employing the M3GNet potential \cite{chen2022universal} for energy calculations.

In recent years, diffusion-based algorithms have emerged as a promising approach offering greater flexibility and efficiency in the generation of novel crystal structures. One notable advancement is the Crystal Diffusion Variational Autoencoder (CDVAE) \cite{xie2021crystal}, which focuses on ab initio crystal generation. This approach employs a Variational Autoencoder (VAE) framework, where crystal compositions are randomly sampled, and the model generates atomic positions and periodic lattice parameters in a denoising diffusion process. By incorporating physical inductive biases, CDVAE ensures invariances such as translation, rotation, and periodic boundary conditions, enabling the generation of stable and realistic crystal structures.
Expanding on this, Cond-CDVAE \cite{luo2024deep} extends CDVAE by leveraging a Noise Conditional Score Network (NCSN) diffusion model to incorporate atomic coordinates and external conditions, such as pressure, into the generation process. Another significant development, DiffCSP \cite{jiao2023crystal}, adopts fractional coordinates instead of multi-graph modeling, allowing simultaneous generation of lattice and atomic coordinates. DiffCSP employs a periodic-E(3)-equivariant denoising model, further refining the prediction process. Building on this, EquiCSP \cite{linequivariant} integrates full E(3)-equivariance into the DiffCSP framework, leveraging periodic graph symmetry and the Denoising Diffusion Probabilistic Model (DDPM) to enhance lattice permutation equivariance and ensure accurate predictions. Another innovative CSP algorithm is ShotgunCSP \cite{chang2024shotgun}, which offers a highly efficient alternative to conventional CSP methods by leveraging a single-shot screening framework. In ShotgunCSP, stable or metastable crystal structures are predicted by identifying the global or local minima of the energy surface within a broad configuration space, without requiring repeated first-principles calculations. However, it may suffer from the limited availability of known structures for a query composition.

Template-based CSP approaches, such as those in TCSP 1.0 \cite{wei2022tcsp} and CSPML \cite{kusaba2022crystal}, have also demonstrated significant potential by narrowing the search space through the use of structural templates, thereby improving computational efficiency. A recent benchmark study of CSP algorithms in CSPBenchmark \cite{wei2024cspbench} showed that template-based CSP algorithms achieved very competitive performance. This is due to the fact that a majority of known crystal structures can be categoried into a few thousands of structure prototypes \cite{hicks2021aflow, zhao2018bias,sapuan2002prototype}. 
CSPML \cite{kusaba2022crystal}, for instance, introduces a machine-learning-based framework that utilizes metric learning to predict isomorphic crystal structures. By training a binary classifier on a dataset of known crystal structures, CSPML achieves high accuracy in identifying structural similarity between different compositions, which allows it to efficiently select template structures from a crystal database for a given query composition. Then it applies element substitution to generate candidate structures, requiring only local relaxation calculations without extensive ab initio simulations.

In this paper, we present TCSP 2.0, an advanced template-based CSP algorithm that integrates more accurate oxidation state prediction models based on the modern deep learning BERTOS model \cite{BERTOS} and sophisticated statistical methods to incorporate chemical knowledge into the ranking model of the TCSP algorithm including element embedding-based similarity metrics, majority space group selection, and CHGNet-based structural relaxation, ensuring improved template selection and structure stability. Through comparative benchmark study across 180 test structures, we evaluate TCSP 2.0 against CSPML and EquiCSP using two primary metrics: space group success rate and StructureMatcher success rate for structural similarity. TCSP 2.0 achieves 78.33\% accuracy in StructureMatcher rate and 83.89\% in space group matching for top-5 predictions, significantly outperforming CSPML and EquiCSP. The Consensus success rate, combining correct structure and space group predictions, reaches 75.00\%, highlighting its superior accuracy and reliability for practical materials discovery.

\begin{figure}[ht]
  \centering
  \includegraphics[width=0.43\linewidth]{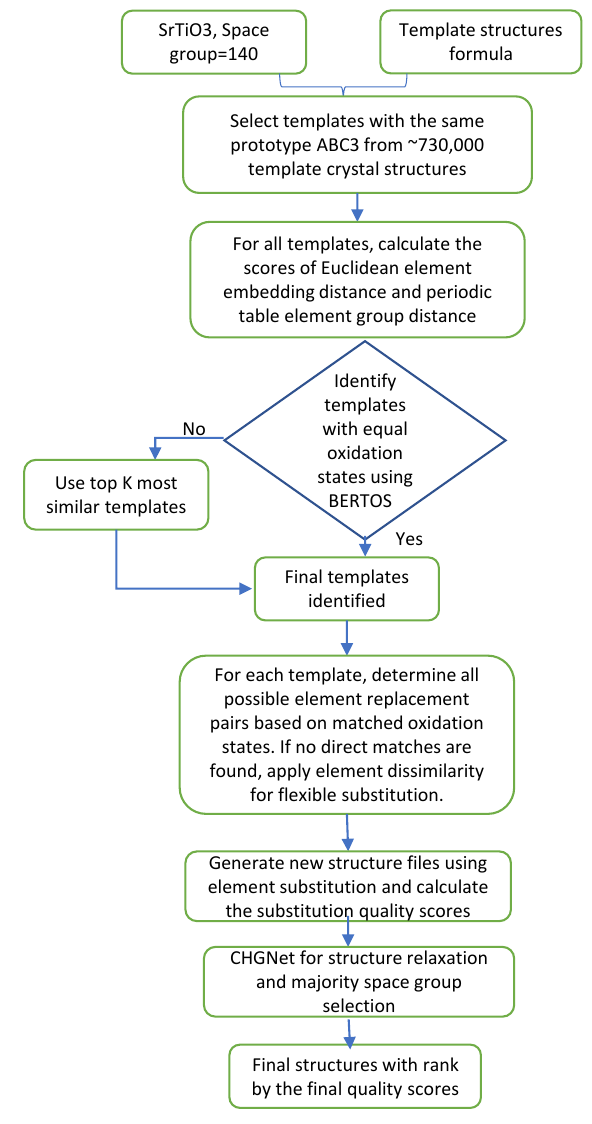}
  \caption{Workflow of TCSP 2.0; The process begins by selecting template structures from the template database based on the prototype. Element substitutions are determined using oxidation state matching and chemical similarity metrics. New structures are then generated and undergo CHGNet-based relaxation and space group determination via majority voting. Finally, the candidate structures are ranked based on their quality scores to identify the most potential candidates.} 
  \label{fig:flowchart}
\end{figure}

\section{Method}
\label{sec:headings}
\subsection{Overview of TCSP algorithm for crystal structure prediction}
TCSP \cite{wei2022tcsp} is a template-based CSP algorithm designed to generate new materials using experimentally known crystal structures as structural templates. Instead of performing expensive global optimization as done in ab initio CSP algorithms or exhaustive enumeration of substitution elements over the template, TCSP efficiently explores the chemical space by substituting elements into carefully selected template structures while preserving their fundamental atomic arrangements. 

As shown in Figure \ref{fig:flowchart}, the structure prediction process of TCSP begins with a database of experimentally validated crystal structures that serve as templates for candidate structure generation. Given a target composition, TCSP first identifies template candidates that share the same prototype from the template set. 
To further refine the selection, all candidates are ranked 
and the top n candidates with the lowest distance scores are chosen. 

Once the template is selected, TCSP applies elemental substitution while strictly enforcing oxidation state matching, meaning that only substitutions are conducted only for element pairs with identical oxidation states in the query composition and in the original template. In this way, all chemically compatible element replacement pairs are explored to generate new structures based on the selected templates while maintaining atomic coordination environments and symmetry. 

Despite its efficiency and ability to generate valid structures, our TCSP 1.0 has several key limitations, which are addressed in this proposed TCSP 2.0.

\paragraph{Limited Accuracy in Oxidation State (OS) Prediction:} TCSP 1.0 relied on the pymatgen oxidation state prediction module, which has an accuracy of less than 15\%. This often leads to incorrect oxidation state assignments, resulting in invalid substitutions and a reduced success rate in generating stable materials. In contrast, TCSP 2.0 integrates BERTOS, a deep learning-based model that achieves 96.82\% OS prediction accuracy across all elemental oxidation states and 97.61\% average accuracy for oxide compounds, significantly improving the reliability of oxidation state assignments and the overall CSP performance as shown in Table \ref{tab:ablation_study} (4th row).

\paragraph{Limited Flexibility in Elemental Replacement:} Without a systematic approach to identifying chemically meaningful element substitutions, TCSP 1.0 struggled to explore alternative compositions effectively, constraining its ability to generate diverse and stable materials. This is due to it being entirely dependent on the Element Mover Distance (ElMD) \cite{hargreaves2020earth} scores to select template structures, which cannot capture the chemical compatibility in the structural atomic local environment. 

\paragraph{Lack of Structural Relaxation:} Generated structures by TCSP 1.0 were directly derived from template-based substitution without further optimization, often resulting in metastable or structurally unrealistic configurations.

These limitations motivated the improvements in TCSP 2.0, which widely expands the template database, introduces more robust similarity metrics, element dissimilarity-based substitution, and CHGNet-based structure relaxation to enhance both the quality and stability of the predicted materials. The whole architecture of the TCSP 2.0 algorithm is shown in Figure \ref{fig:flowchart}.

\subsection{Dataset Expansion for TCSP 2.0} In TCSP 1.0, only the Materials Project (MP) database \cite{jain2013commentary} was used as the template source. While MP provides a well-curated and widely used dataset of inorganic materials, its scope is limited compared to the broader landscape of available crystal structure databases.

We expanded the dataset to include 731,293 crystal structures drawn from multiple sources, incorporating additional databases to enhance the prediction accuracy and structural diversity for TCSP 2.0. These sources include the Materials Project (MP) database, the Materials Cloud database (both 2D and 3D) \cite{materialscloud}, the Computational 2D Materials Database (C2DB) \cite{gjerding2021recent}, and the Graph Networks for Materials Science database (GNoME) \cite{merchant2023scaling}. 

\paragraph{MP:} The MP database is one of the most widely used crystal material repositories for computational materials science, containing over 150,000 inorganic crystal structures along with computed properties such as formation energy, band structure, and elastic constants. MP is particularly valuable because of its high-quality first-principles calculations and standardization, making it a strong foundation for crystal structure generation and prediction.

\paragraph{Materials Cloud:} The Materials Cloud database \cite{talirz2020materials} is a platform that aggregates structures from various high-throughput materials science studies. It contains both bulk (3D) and two-dimensional (2D) materials, broadening the scope of possible template structures. Incorporating 2D materials into TCSP 2.0 allows for the exploration of layered materials and potential novel materials beyond conventional 3D structures.

\paragraph{C2DB:} The C2DB \cite{gjerding2021recent} database focuses specifically on two-dimensional materials, providing 16,789 unique structures along with computed properties such as electronic band gaps, magnetic ordering, and mechanical stability. By integrating C2DB, TCSP 2.0 gains access to a diverse range of 2D materials, which are essential for applications in nanoelectronics, catalysis, and energy storage \cite{zhu2018structural, pomerantseva2019energy, yun2020layered}. The inclusion of C2DB improves the ability of TCSP to generate 2D materials with tunable properties.

\paragraph{GNoME:} GNoME \cite{merchant2023scaling} is a recently developed large-scale dataset generated using graph neural networks to accelerate inorganic crystal discovery. It expands known stable materials by an order of magnitude, identifying 2.2 million new structures below the convex hull and building on 48,000 stable crystals identified in continuing studies, with 736 experimentally confirmed. The advantage of incorporating GNoME into TCSP 2.0 lies in its ability to provide a larger set of realistic but previously unexplored structures, improving both coverage and diversity in material generation.

By integrating these diverse databases, TCSP 2.0 ensures a more comprehensive representation of crystal structures, facilitating the discovery of novel materials while maintaining accuracy and chemical validity. The expanded dataset provides better generalization capabilities, allowing for more reliable template selection and improved elemental substitutions across a wider range of materials.

\subsection{CHGNet Integration in TCSP 2.0} 
In recent years, numerous machine learning (ML) interatomic potentials have been developed for efficient and accurate structure optimization, providing an alternative to computationally expensive ab initio methods. Models such as MEGnet\cite{chen2019graph}, M3GNet \cite{chen2022universal}, and CHGNet \cite{deng2023chgnet} have demonstrated remarkable capabilities in predicting atomic forces, energy landscapes, and structural stability with near-DFT accuracy. These ML potentials leverage large datasets of first-principles calculations to learn interatomic interactions, allowing for fast and scalable relaxation of crystal structures across diverse material families. Among these, CHGNet stands out as a pretrained universal neural network potential that incorporates charge-informed atomistic modeling, making it particularly well-suited for materials science applications.

To enhance the structural stability and energy evaluation of predicted materials, TCSP 2.0 integrates CHGNet, a pretrained universal neural network potential designed for charge-informed atomistic modeling. CHGNet has demonstrated exceptional accuracy in structure relaxation and energy prediction, rivaling first-principles calculations while maintaining high computational efficiency. By leveraging its advanced graph neural network architecture, CHGNet enables fast and precise relaxation of complex crystal structures, ensuring that TCSP-generated materials are physically meaningful and experimentally viable.
CHGNet achieves state-of-the-art accuracy in predicting total energies, atomic forces, and charge distributions, closely matching results from density functional theory (DFT). It significantly outperforms traditional empirical potentials and provides a reliable alternative to computationally expensive ab initio methods.

\subsection{Element Dissimilarity for Flexible Replacement}
In TCSP 1.0, candidate selection was strictly limited to elements with matching oxidation states, which significantly constrained the diversity of generated materials. In TCSP 2.0, we introduce an element dissimilarity-based replacement strategy to overcome this limitation. Instead of discarding candidates that do not share the same oxidation states, we use a precomputed element dissimilarity matrix to identify chemically similar replacements, ensuring that the substitution remains meaningful.

When no candidate template formula shares the same oxidation states as the query composition, we first represent both the query and candidate formulas as composition vectors. The elements in each formula are ranked based on their atomic fractions, and unique composition groups are identified. Within each group, we systematically search for replacement pairs that minimize the overall element dissimilarity. If multiple substitutions exist, we evaluate all permutations and select the configuration with the lowest total dissimilarity score.

This new approach significantly improves the flexibility of TCSP 2.0, enabling the discovery of more diverse and stable materials. Unlike TCSP 1.0, which often failed to generate viable candidates due to strict oxidation state constraints, the new method ensures that substitutions remain chemically reasonable and structurally coherent. By allowing chemically similar elements to replace one another, TCSP 2.0 expands the search space and enhances the likelihood of discovering new functional materials.

\subsection{Enhancing Oxidation State Prediction in Template Search Processes Using BERTOS} 

In our previous TCSP approach, oxidation state prediction relied on the \texttt{pymatgen} package \cite{pymatgen}, which has limited accuracy of < 15\%. To address this, we incorporated the BERTOS model \cite{BERTOS}, a deep learning-based transformer architecture designed for composition-based oxidation state prediction. BERTOS demonstrates exceptional accuracy, with 96.82\% precision across all elemental oxidation states on a curated Inorganic Crystal Structure Database (ICSD) dataset, and 97.61\% accuracy for oxide compounds. Unlike traditional heuristic-based methodologies, BERTOS leverages a sophisticated deep learning transformer architecture to predict oxidation states directly from chemical compositions. This approach circumvents the conventional reliance on structural information, introducing a more generalized and data-driven predictive framework.
Integrating BERTOS into our TCSP 2.0 framework significantly enhances the accuracy and reliability of oxidation state predictions. This strategic incorporation enables more robust charge-neutrality verification and refines template matching strategies during crystal structure search processes, potentially improving the overall precision of structural predictions.

\subsection{Distance Metrics for finding top-k template candidates}
To improve the selection of template candidates and elemental substitutions in TCSP 2.0, we incorporate a combination of element embedding distance and periodic table group distance. These metrics address the limitations of previous approaches by capturing both chemical similarity and structural constraints, ensuring that substitutions lead to stable and meaningful materials. The element embedding distance provides a data-driven measure of elemental similarity, while the periodic table group distance enforces constraints based on periodic trends. By integrating these two metrics, TCSP 2.0 refines the search process, allowing for better template selection and more chemically coherent substitutions, ultimately enhancing the stability and diversity of predicted materials.

\paragraph{Euclidean Element Embedding Distance:} 
In TCSP 2.0, element embedding distance is leveraged to enhance both template selection and elemental substitution, ensuring that generated structures maintain chemical and structural validity. Instead of relying solely on the Element Mover’s Distance (ElMD), we incorporate element embeddings that encode learned representations of elemental properties. The Euclidean embedding distance quantifies the similarity between elements in this high-dimensional space, enabling a more refined and chemically meaningful selection of candidate structures.

To compute the embedding distance, we begin by sorting lists of elements based on their electronegativity ($X$) and atomic number. In the context of our computation, \texttt{comp1} refers to a specific composition dataset, and \texttt{comp1.elements} denotes the list of elements within this dataset. The sorting operation is executed using Python's \texttt{sorted} function, which arranges elements according to a specified key function. In our case, the key function is defined as $\lambda \ e: (e.X, e.\text{number})$, where \texttt{e} represents each element in \texttt{comp1.elements}. Here, \texttt{e.X} signifies the electronegativity of the element \texttt{e}, and \texttt{e.number} represents its atomic number. This sorting process ensures that elements are consistently ordered across different compositions, facilitating accurate comparisons in our embedding distance calculations.

\begin{equation}
\text{elements} = \text{sorted}(\text{comp}_1.\text{elements}, \text{key}=\lambda \ e: (e.X, e.\text{number}))
\end{equation}

During the search process, the top-$ k $ template candidates are ranked based on their embedding distance scores relative to the target composition. This ensures that during the search process, the top-$ k $ template candidates are ranked based on their embedding distance scores to the target composition, ensuring that the selected templates align more closely with the desired material properties.
Quantifying the similarity between chemical elements requires sophisticated distance metrics in materials science. We propose a comprehensive approach that combines embedding space distance with periodic table characteristics. The Euclidean embedding distance between two elements \( e_1 \) and \( e_2 \) is defined as:

\begin{equation}
d_{\text{embedding}}(e_1, e_2) = \sqrt{\sum_{k=1}^{n} (v_{1,k} - v_{2,k})^2}
\tag{2}
\end{equation}

where \( n \) represents the dimensionality of the embedding space, and \( v_{1,k} \) and \( v_{2,k} \) are the \( k \)-th components of the embedding vectors for elements \( e_1 \) and \( e_2 \), respectively.

\paragraph{Periodic Table Element Group Distance:} 
To further refine elemental substitution, we introduce a periodic table group distance as an additional constraint. This metric accounts for the structural and chemical relationships dictated by periodic trends, ensuring that substitutions occur within chemically meaningful groups. Elements within the same periodic group share the same valence electron configurations and bonding characteristics, making them more suitable replacements.
Elements within the same group exhibit the following common properties:

\begin{itemize}
\item \textbf{Electronic Configuration:} They have the same number of valence electrons, leading to similar bonding behavior.
\item \textbf{Chemical Reactivity:} They form similar compounds and exhibit comparable reaction patterns.
\item \textbf{Periodic Trends:} They follow similar trends in atomic radius, ionization energy, and electronegativity.
\item \textbf{Oxidation States:} They tend to exhibit consistent oxidation states, influencing their role in chemical reactions.
\end{itemize}
For example, alkali metals (Group 1) such as Li, Na, and K are highly reactive, forming +1 oxidation states. Halogens (Group 17), including F, Cl, and Br, readily gain electrons to form -1 oxidation states. Noble gases (Group 18) are chemically inert due to their stable electron configurations.

During candidate generation, we compute both the embedding-based similarity and the periodic table group penalty, prioritizing substitutions that minimize the total distance score. To formalize this constraint, we define the group distance penalty as follows:

\begin{equation}
\Delta_g(e_1, e_2) = 
\begin{cases} 
1, & \text{if } \text{group}(e_1) \neq \text{group}(e_2), \\
0, & \text{otherwise}.
\end{cases}
\tag{3}
\end{equation}

The total element replacement score \( S \) combines both embedding and group distance penalty:

\begin{equation}
S(e_1, e_2) = d_{\text{embedding}}(e_1, e_2) + \Delta_g(e_1, e_2)
\tag{4}
\end{equation}

By incorporating both element similarity (via element embeddings) and structural constraints (via periodic trends), this approach provides a more comprehensive measure of element substitutability. This ensures that substitutions remain chemically reasonable and structurally coherent, leading to more stable and experimentally viable materials in TCSP 2.0.

\subsection{Majority Voting based Space Group Selection:}
In crystal structure prediction, identifying the most probable structural configuration extends beyond minimal distance metrics. Our approach implements a novel majority voting mechanism for space group selection, which systematically evaluates candidate structures based on their group consistency. If a dominant space group appears in more than 60\% of the top-$ k $ predicted structures, we apply a penalty to minority group structures.
Specifically, the algorithm identifies the most frequently occurring space group among candidate structures. Structures deviating from this dominant group receive an additional score penalty, effectively prioritizing configurations with collective structural coherence. This strategy leverages the statistical patterns inherent in crystal structure predictions, enhancing the reliability of our selection process by incorporating ensemble-based reasoning. While the majority space group may not invariably represent the true crystal structure, our empirical investigations demonstrate that this heuristic approach systematically improves overall prediction performance.

\FloatBarrier

\section{Results and discussion}
To evaluate the performance of our TCSP 2.0, we used the 180 benchmark test materials from the CSPBenchmark \cite{wei2024cspbench}. These materials were chosen to comprehensively assess an algorithm’s ability to predict structures that closely match the target structures, providing a robust evaluation of its predictive accuracy and reliability.

\subsection{Comparison results of TCSP 2.0, EquiCSP, and CSPML algorithms}
We compared the performance of TCSP 2.0 with two recent crystal structure prediction algorithms: CSPML, a template-based CSP algorithm, and EquiCSP \cite{kusaba2022crystal, linequivariant}, a diffusion-based CSP algorithm. CSPML approaches crystal structure prediction through metric learning and operates by selecting template structures from the MP database and applying element substitution, mirroring traditional materials discovery protocols in a machine learning framework.
EquiCSP represents a different approach, implementing an equivariant diffusion-based generative model that addresses fundamental symmetry considerations in crystal structure prediction. The algorithm uniquely maintains lattice permutation equivariance and periodic translation equivariance throughout both training and inference processes. This comprehensive treatment of crystal symmetries enables EquiCSP to generate high-quality structures while demonstrating faster convergence during training compared to previous methods. For a fair comparison, we removed 180 benchmark test data from the training set for all algorithms. 

\begin{figure}[ht]
  \centering
  \includegraphics[width=0.87\linewidth]{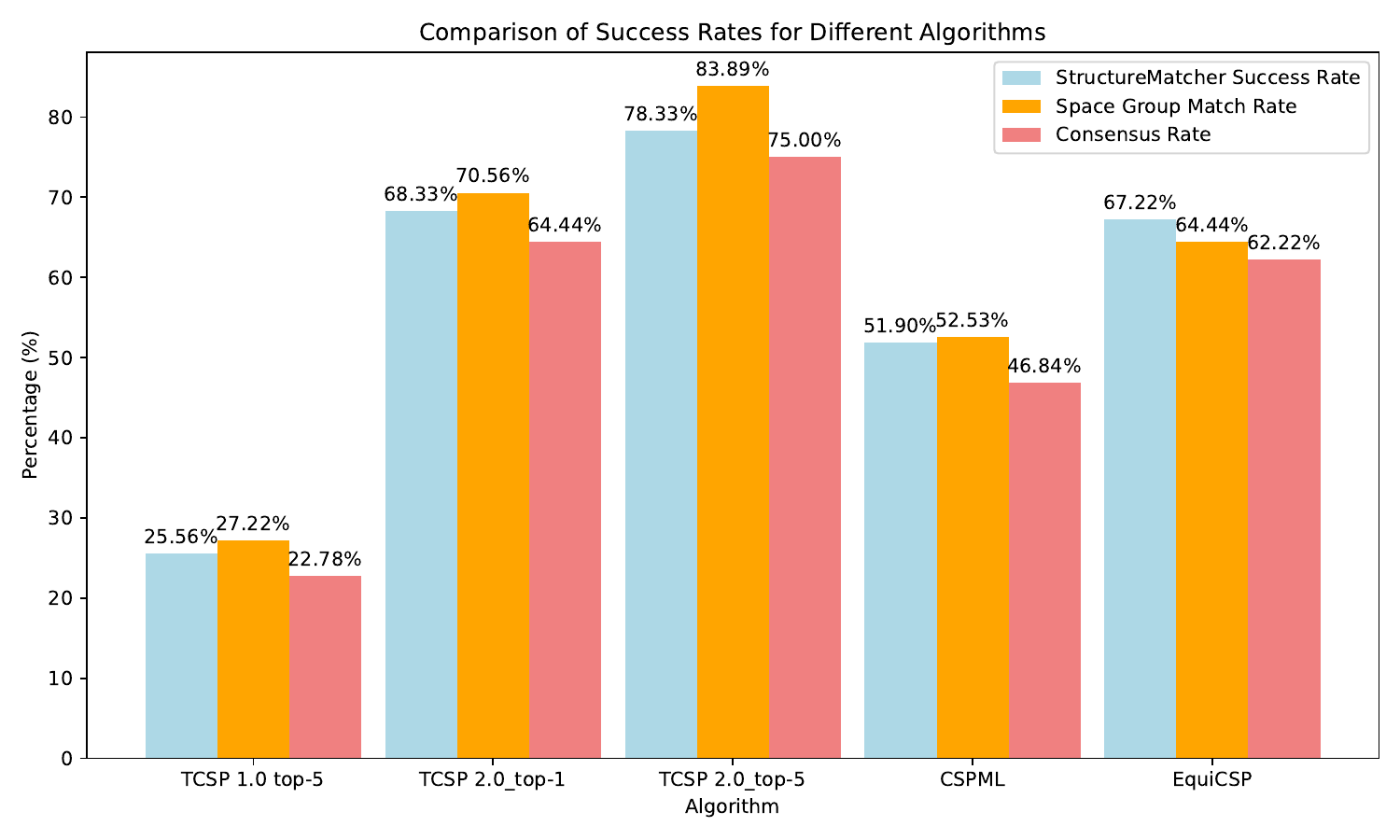}
  \caption{Performance of Success Rates for StructureMatcher, Space Group Match Rates, and Consensus Rates Across TCSP 1.0, TCSP 2.0, CSPML, EquiCSP. The Consensus Rate highlights the combined effectiveness of both StructureMatcher and Space Group success for each algorithm. } 
  \label{fig:validation}
\end{figure}

\FloatBarrier

\paragraph{StructureMatcher and Space Group Success Rates}
First, we utilized the pymatgen package \cite{pymatgen} to evaluate the prediction accuracy of TCSP 2.0, CSPML, and EquiCSP across the 180 test cases. Two primary metrics were employed: the space group success rate, which measures the algorithms' ability to correctly predict the space group symmetry of crystal structures, and the structure matcher comparison, which assesses the structural similarity between predicted and ground truth structures. In our evaluation framework, we consider two types of predictions: top-1 and top-5. We define top-1 as the single highest-scoring predicted structure for a given formula. Meanwhile, the top-5 predictions come from setting the top-$ k $ to 5, meaning the algorithm can return up to five predicted structures for each formula. A prediction is considered successful if the correct structure appears among those five. 
Figure \ref{fig:validation} illustrates the performance comparison across these metrics. TCSP 2.0 exhibits a substantial improvement over TCSP 1.0, CSPML, and EquiCSP, particularly in top-5 predictions. TCSP 2.0’s StructureMatcher success rate improves from 68.33\% in top-1 to 78.33\% in top-5, while its Space Group match rate increases from 70.56\% to 83.89\%, highlighting the advantage of allowing multiple candidates. The Consensus Rate follows a similar trend, rising from 64.44\% in top-1 to 75.00\% in top-5, demonstrating the reliability of TCSP 2.0 in generating physically meaningful structures.

Comparatively, EquiCSP achieves 67.22\% StructureMatcher accuracy, 64.44\% Space Group match rate, and 62.22\% Consensus Rate, showing strong but slightly lower performance than TCSP 2.0. CSPML performs more modestly, with 51.90\% for StructureMatcher, 52.53\% for Space Group matching, and 46.84\% for Consensus Rate, indicating that its generative approach struggles with structural accuracy. TCSP 1.0, constrained by its rigid oxidation state matching and lack of structural relaxation, achieves significantly lower success rates, with 25.56\% for StructureMatcher, 27.22\% for Space Group match rate, and 22.78\% for Consensus Rate.

Overall, these results emphasize TCSP 2.0’s superior performance in crystal structure prediction, particularly when leveraging the top-5 predictions. Its integration of BERTOS for oxidation state prediction, element dissimilarity-based substitution, and CHGNet relaxation enables a more effective and reliable framework for materials discovery.

\subsection{Ablation study}
To systematically evaluate the performance improvements introduced in TCSP 2.0, we conducted an ablation study focusing on its most critical components. We compared the original TCSP 1.0, which relies solely on ElMD (Element’s Mover Distance) \cite{hargreaves2020earth}, against three progressively enhanced configurations: one incorporating BERTOS for oxidation state prediction, another adding element embedding and group distance, and the full TCSP 2.0 system, which further integrates ML potential-based structure relaxation and space group selection. For each configuration, we assessed performance using three key metrics: the StructureMatcher success rate, which evaluates structural similarity; the Space Group success rate, which measures symmetry prediction accuracy; and the Consensus rate, which reflects overall prediction consistency. The comparative results are summarized in Table \ref{tab:ablation_study}.

In TCSP 1.0, which relies exclusively on ElMD for template matching, the StructureMatcher success rate is 25.6\%, the Space Group success rate is 27.2\%, and the Consensus rate is 22.8\%. Incorporating BERTOS for oxidation state prediction leads to substantial improvements, raising the StructureMatcher success rate to 48.3\%, the Space Group success rate to 51.7\%, and the Consensus rate to 45.6\%. Moreover, this enhancement significantly increases the number of successful predictions from 88 to 169 cases. The key reason for this improvement is that oxidation state prediction in TCSP 1.0 relied on pymatgen, which has a low accuracy of approximately 15\%, limiting the number of successfully predicted structures. Since oxidation states must be correctly assigned before template matching can proceed, the improved accuracy with BERTOS allows for more valid predictions, enabling a larger number of structures to advance to subsequent steps.

Further improvements are observed when replacing ElMD with element embedding and group distance, while also incorporating element dissimilarity (ED) for unmatched oxidation states, all while maintaining BERTOS for oxidation state prediction. With this configuration, the StructureMatcher success rate increases to 55.7\%, the Space Group success rate improves to 59.8\%, and the Consensus rate reaches 54.0\%. This demonstrates the enhanced effectiveness of our element similarity measures in guiding better template selection and chemically valid substitutions. By leveraging element embeddings and group distance, this approach captures more nuanced elemental relationships than ElMD, allowing for more accurate template matching. Additionally, introducing ED for unmatched oxidation states refines the substitution process by mitigating errors caused by incorrect oxidation states, further improving structure generation accuracy.

The full TCSP 2.0 system, which incorporates Space Group Selection in addition to the improvements above, delivers the best results. In this configuration, the StructureMatcher success rate reaches 68.3\%, representing a 167\% improvement over TCSP 1.0. The Space Group success rate also rises significantly to 70.6\%, while the Consensus rate improves to 64.4\%, highlighting the consistency between structural similarity and symmetry predictions. The addition of Space Group Selection plays a critical role in this performance boost, as it refines the final structure prediction by selecting the most probable space group based on an ensemble of candidate structures. This process significantly reduces errors introduced by incorrect initial assignments and enhances the overall robustness of the framework.

Overall, these results demonstrate that TCSP 2.0 offers a significantly more reliable and effective framework for crystal structure prediction compared to TCSP 1.0. Each enhancement systematically improves different aspects of the prediction process, ultimately leading to higher success rates across all key metrics. These findings underscore the importance of integrating advanced element similarity metrics, robust oxidation state prediction, and space group optimization in the design of next-generation crystal structure prediction models.

\begin{table}[ht]
\caption{Ablation study results showing the effect of various components on TCSP 2.0 performance. ElMD: Element's Mover Distance; ED: element dissimilarity for unmatched OS; Consensus rate: the proportion of predictions that achieve both StructureMatcher and Space Group success.}
\centering
\begin{tabular}{|lllll|l|l|l|}
\hline
\multicolumn{5}{|c|}{\textbf{Ablation Settings}} & \multicolumn{1}{c|}{\multirow{2}{*}{\textbf{\begin{tabular}[c]{@{}c@{}}StructureMatcher\\ success rate\end{tabular}}}} & \multicolumn{1}{c|}{\multirow{2}{*}{\textbf{\begin{tabular}[c]{@{}c@{}}Space Group \\ success rate\end{tabular}}}} & \multicolumn{1}{c|}{\multirow{2}{*}{\textbf{\begin{tabular}[c]{@{}c@{}}Consensus\\ rate\end{tabular}}}} \\
\cline{1-5}
\multicolumn{1}{|l|}{ElMD} & \multicolumn{1}{l|}{\begin{tabular}[c]{@{}l@{}}Element embedding \\ and group distance\end{tabular}} & \multicolumn{1}{l|}{BERTOS} & \multicolumn{1}{l|}{ED} & \begin{tabular}[c]{@{}l@{}}Space Group \\ Selection\end{tabular} & \multicolumn{1}{c|}{} & \multicolumn{1}{c|}{} & \\
\hline
\multicolumn{1}{|c|}{\ding{51}} & \multicolumn{1}{c|}{\ding{55}} & \multicolumn{1}{c|}{\ding{55}} & \multicolumn{1}{c|}{\ding{55}} & \multicolumn{1}{c|}{\ding{55}} & 25.6\% & 27.2\% & 22.8\% \\
\hline
\multicolumn{1}{|c|}{\ding{51}} & \multicolumn{1}{c|}{\ding{55}} & \multicolumn{1}{c|}{\ding{51}} & \multicolumn{1}{c|}{\ding{55}} & \multicolumn{1}{c|}{\ding{55}} & 48.3\% & 51.7\% & 45.6\% \\
\hline

\multicolumn{1}{|c|}{\ding{55}} & \multicolumn{1}{c|}{\ding{51}} & \multicolumn{1}{c|}{\ding{51}} & \multicolumn{1}{c|}{\ding{51}} & \multicolumn{1}{c|}{\ding{55}} & 55.7\% & 59.8\% & 54.0\% \\
\hline
\multicolumn{1}{|c|}{\ding{55}} & \multicolumn{1}{c|}{\ding{51}} & \multicolumn{1}{c|}{\ding{51}} & \multicolumn{1}{c|}{\ding{51}} & \multicolumn{1}{c|}{\ding{51}} & 68.3\% & 70.6\% & 64.4\% \\
\hline
\end{tabular}
\label{tab:ablation_study}
\end{table}

\begin{figure}[!htb]
  \centering
  \includegraphics[width=0.8\linewidth]{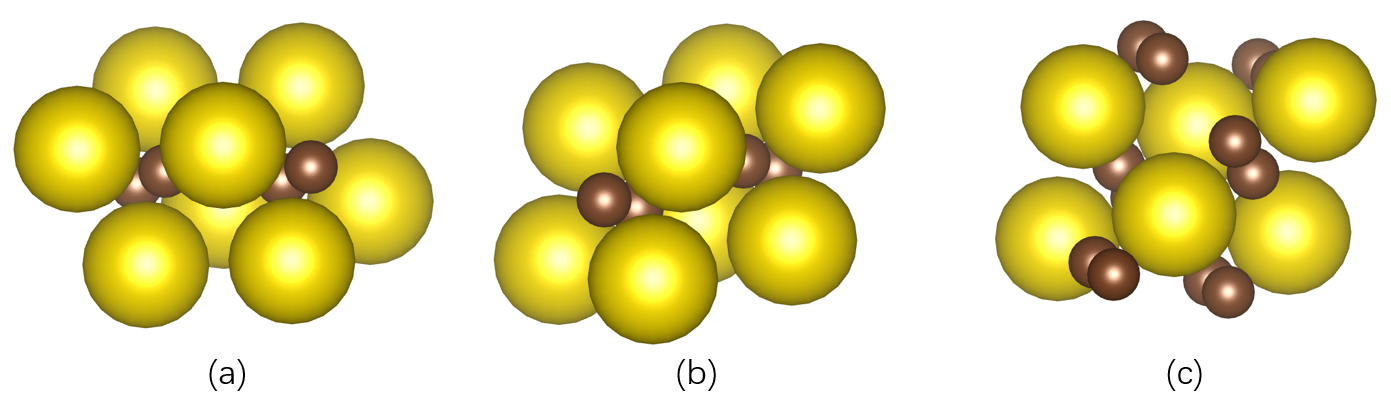}
  \caption{Comparison of TCSP 1.0 and TCSP 2.0 predictions of PrC2. (a) Ground truth crystal structure. (b) Predicted structure using TCSP 2.0, closely resembling the ground truth. (c) Predicted structure using TCSP 1.0, which fails to capture the correct atomic arrangement.}
  \label{fig:PrC2}
\end{figure}

\begin{figure}[!htb]
  \centering
  \includegraphics[width=0.8\linewidth]{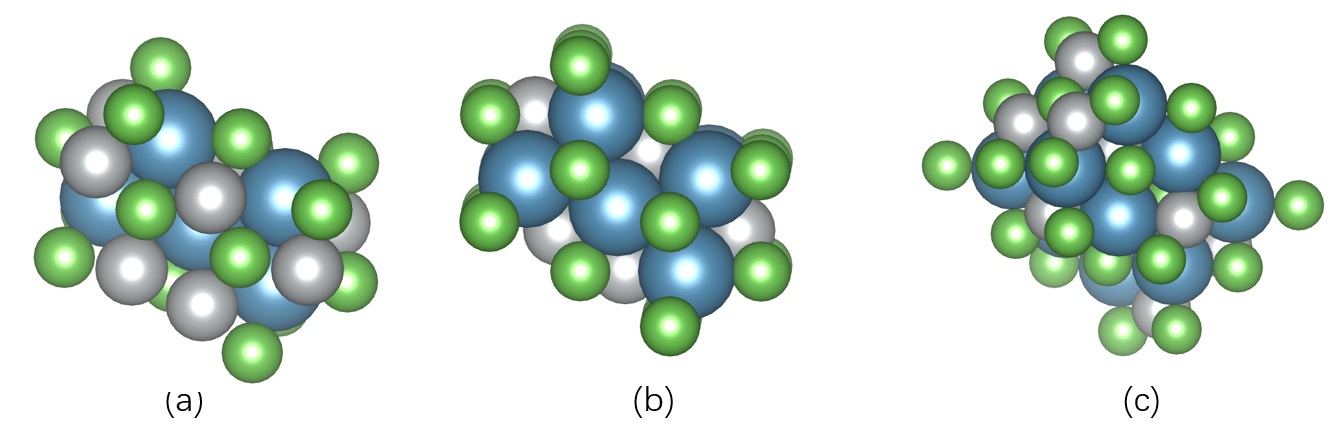}
  \caption{Comparison of TCSP 1.0 and TCSP 2.0 predictions of Ca3Ag3As3. (a) Ground truth crystal structure. (b) Predicted structure using TCSP 2.0, closely resembling the ground truth. (c) Predicted structure using TCSP 1.0, which fails to capture the correct atomic arrangement.}
  \label{fig:Ca3Ag3As3}
\end{figure}

\begin{figure}[!htb]
  \includegraphics[width=0.9\linewidth]{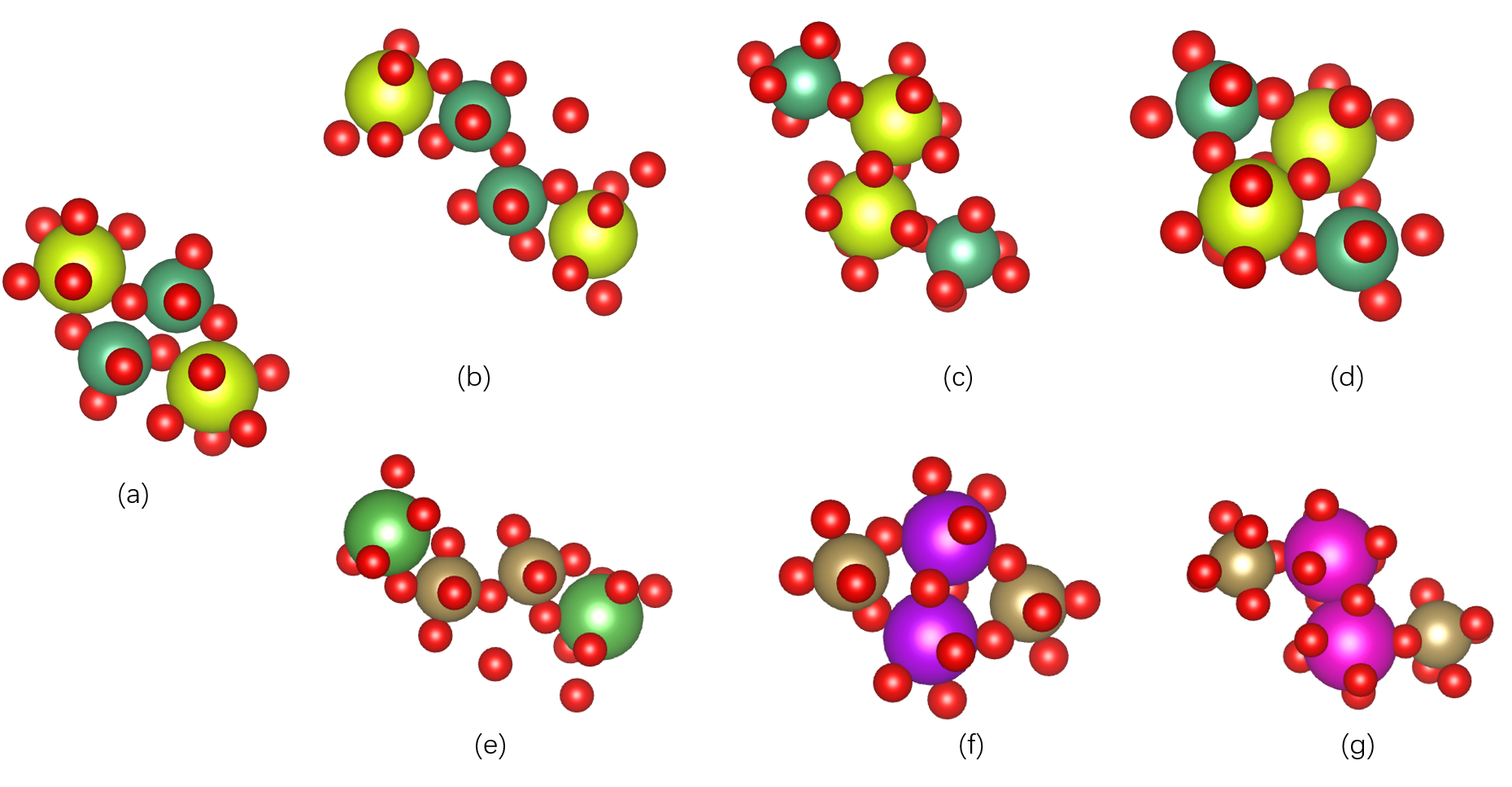}
\caption{Case study on the majority space group selection mechanism in TCSP 2.0 for Ce\textsubscript{2}Nb\textsubscript{2}O\textsubscript{8}. (a) Ground truth structure with space group 15. (b) Top-ranked prediction (La\textsubscript{2}Ta\textsubscript{2}O\textsubscript{8}, mp-3998), but with incorrect space group 36. (c-d) Second and third-ranked predictions (Gd\textsubscript{2}Ta\textsubscript{2}O\textsubscript{8}, mp-5575 and Eu\textsubscript{2}Ta\textsubscript{2}O\textsubscript{8}, mp-4271) correctly match space group 15. Without majority space group selection, the incorrect prediction (b) would have been chosen. (e-g) Template structures used for predictions (b-d).}
  \label{fig:Ce2Nb2O8}
\end{figure}

\begin{figure}[!htb]
  \centering
  \includegraphics[width=0.9\linewidth]{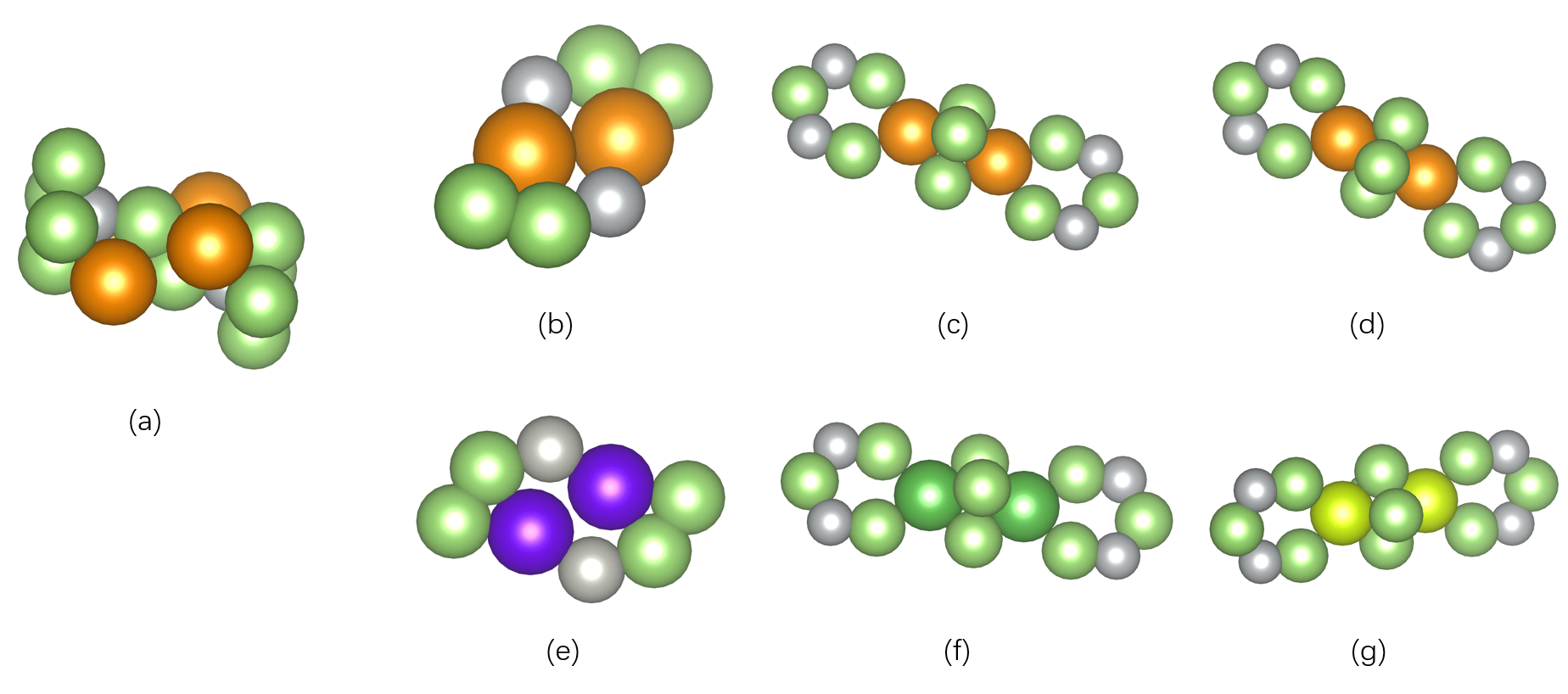}
  \caption{Case study on the majority space group selection mechanism in TCSP 2.0 for Nd\textsubscript{2}Ga\textsubscript{4}Ni\textsubscript{2}. (a) Ground truth structure with space group 65. (b) Top-ranked predicted structure (Tb\textsubscript{2}Ga\textsubscript{4}Pd\textsubscript{2}, mp-972423) with a score of 0.81, but incorrectly assigned to space group 63. (c) Second-ranked predicted structure (La\textsubscript{2}Ga\textsubscript{4}Ni\textsubscript{2}, MC3D) with a score of 0.81, correctly identifying space group 65. (d) Third-ranked predicted structure (Ce\textsubscript{2}Ga\textsubscript{4}Ni\textsubscript{2}, mp-1025446) with a score of 0.84, also correctly identifying space group 65. Without majority space group selection, the incorrect prediction (b) would have been chosen. The majority voting mechanism prioritizes (c) and (d), improving prediction accuracy. (e-g) Template structures used for generating the predictions (b-d).}
  \label{fig:Nd2Ga4Ni2}
\end{figure}

To further assess the performance improvements in TCSP 2.0, we compare its predictions against TCSP 1.0 for multiple material systems. Figures \ref{fig:PrC2} and \ref{fig:Ca3Ag3As3} illustrate two cases—PrC\textsubscript{2} and Ca\textsubscript{3}Ag\textsubscript{3}As\textsubscript{3}.

In TCSP 1.0, which relies solely on ElMD for template matching (as in the first row of the ablation study), the predictions often fail to capture the correct atomic arrangement due to the lack of oxidation state prediction and advanced template selection mechanisms. Figure \ref{fig:PrC2}(a) shows the ground truth crystal structure, while Figure \ref{fig:PrC2}(b) presents the structure predicted by TCSP 2.0, which accurately reproduces the correct coordination environment. In contrast, Figure \ref{fig:PrC2}(c) illustrates the TCSP 1.0 prediction, which fails to reconstruct the proper atomic arrangement, highlighting the limitations of relying solely on ElMD.

The improvements in TCSP 2.0, including BERTOS for oxidation state prediction, enhanced template selection via element embedding and group distance, and CHGNet-based structure relaxation, enable more accurate and reliable predictions, as seen in Figure \ref{fig:PrC2}(b). A similar trend is observed for Ca\textsubscript{3}Ag\textsubscript{3}As\textsubscript{3} (Figure \ref{fig:Ca3Ag3As3}), where TCSP 2.0 successfully predicts a structure closely matching the ground truth, while TCSP 1.0, using only ElMD, fails to generate an accurate prediction.

To further illustrate the effectiveness of the majority space group selection mechanism in TCSP 2.0, we present two case studies: Ce\textsubscript{2}Nb\textsubscript{2}O\textsubscript{8} (Figure \ref{fig:Ce2Nb2O8}) and Nd\textsubscript{2}Ga\textsubscript{4}Ni\textsubscript{2} (Figure \ref{fig:Nd2Ga4Ni2}). These examples demonstrate how majority voting ensures that space group remain consistent across multiple candidate structures, thereby improving the overall accuracy and reliability of structural predictions.
In the case of Ce\textsubscript{2}Nb\textsubscript{2}O\textsubscript{8}, the ground truth structure belongs to space group 15 (Figure \ref{fig:Ce2Nb2O8}(a)). Figures \ref{fig:Ce2Nb2O8}(b-d) show the predicted structures, while Figures \ref{fig:Ce2Nb2O8}(e-g) display the templates used for their generation. The top-ranked predicted structure (Figure \ref{fig:Ce2Nb2O8}(b)) was matched with La\textsubscript{2}Ta\textsubscript{2}O\textsubscript{8} (mp-3998), achieving a score of 0.59, but it belongs to space group 36, which does not match the ground truth. The second-ranked structure (Figure \ref{fig:Ce2Nb2O8}(c)) was matched with Gd\textsubscript{2}Ta\textsubscript{2}O\textsubscript{8} (mp-5575), scoring 0.60, and correctly identifying space group 15. Similarly, the third-ranked structure (Figure \ref{fig:Ce2Nb2O8}(d)) matched Eu\textsubscript{2}Ta\textsubscript{2}O\textsubscript{8} (mp-4271) with a score of 0.61, also correctly identifying space group 15.

Without the majority space group selection, the incorrect prediction (Figure \ref{fig:Ce2Nb2O8}(b)) would have been chosen based on its ranking score, leading to an inaccurate structure. However, by applying the majority voting mechanism, TCSP 2.0 correctly prioritizes predictions with consistent space group assignments (Figures \ref{fig:Ce2Nb2O8}(c) and \ref{fig:Ce2Nb2O8}(d)), significantly enhancing the accuracy and reliability of structure prediction. Figures \ref{fig:Ce2Nb2O8}(e-g) show the templates used to generate these structures, further illustrating the role of template selection in the final predictions.

For Nd\textsubscript{2}Ga\textsubscript{4}Ni\textsubscript{2}, the ground truth structure belongs to space group 65 (Figure \ref{fig:Nd2Ga4Ni2}(a)). Figures \ref{fig:Nd2Ga4Ni2}(b-d) display the predicted structures, while Figures \ref{fig:Nd2Ga4Ni2}(e-g) show the template structures used for their generation. The top-ranked predicted structure (Figure \ref{fig:Nd2Ga4Ni2}(b)) was matched with Tb\textsubscript{2}Ga\textsubscript{4}Pd\textsubscript{2} (mp-972423), achieving a score of 0.81, but it belongs to space group 63, which does not match the ground truth. Meanwhile, the second-ranked (Figure \ref{fig:Nd2Ga4Ni2}(c)) and third-ranked (Figure \ref{fig:Nd2Ga4Ni2}(d)) predictions, matched with La\textsubscript{2}Ga\textsubscript{4}Ni\textsubscript{2} (MC3D) and Ce\textsubscript{2}Ga\textsubscript{4}Ni\textsubscript{2} (mp-1025446), respectively, with scores of 0.81 and 0.84, correctly identified space group 65.

Without the majority space group selection mechanism, the incorrect prediction (Figure \ref{fig:Nd2Ga4Ni2}(b)) would have been chosen, leading to an inaccurate assignment. However, TCSP 2.0 correctly prioritizes predictions that consistently match the correct space group, ensuring greater reliability in structural predictions. Figures \ref{fig:Nd2Ga4Ni2}(e-g) illustrate the templates used in generating these predictions also demonstrate the crucial role of template selection in the final results.

Collectively, these results demonstrate that TCSP 2.0 significantly outperforms TCSP 1.0 by improving oxidation state prediction, template selection, and space group refinement. These enhancements make TCSP 2.0 a robust and reliable framework for crystal structure prediction.

\section{Conclusion}
In this work, we present TCSP 2.0, an advanced template-based crystal structure prediction framework that significantly improves on its predecessor through the integration of modern deep learning techniques and sophisticated statistical methods. The key innovations of our approach include:

The incorporation of the BERTOS model for oxidation state prediction, which achieves 96.82\% precision across elemental oxidation states, substantially improving the accuracy of charge-neutrality verification in the structure prediction process.
A sophisticated distance metric that combines element embedding distances with periodic table group constraints, providing a more nuanced approach to element substitutability in template matching.
A novel majority voting mechanism for space group selection that leverages ensemble information to enhance structural prediction reliability.
The expansion of template sources from the original Materials Project database to include Materials Cloud, C2DB, and GNoME databases, significantly broadening the coverage of potential structural templates.

Evaluations on 180 benchmark materials show TCSP 2.0 achieving 83.89\% space group success rate and 78.33\% structural similarity accuracy for top-5 predictions, significantly outperforming CSPML and EquiCSP, with a 75.00\% consensus rate for combined structure and space group accuracy. In contrast, TCSP 1.0, which relied solely on Element Mover’s Distance (ElMD) and lacked advanced oxidation state prediction and structural relaxation, achieved only 25.56\% StructureMatcher success rate, 27.22\% space group success rate, and a 22.78\% consensus rate, highlighting the improvements introduced in TCSP 2.0. These enhancements make TCSP 2.0 a more accurate and reliable tool for materials discovery, bridging the gap between computational efficiency and predictive accuracy. The open-source implementation of our framework further supports its adoption and extension by the broader materials science community. 

Potential future directions include integrating deep learning methods, such as ROOST \cite{goodall2020predicting}, to enhance the composition-based scoring process. By leveraging ROOST, we can compute a composition score that guides template candidate selection more effectively. Incorporating a learned representation of composition through ROOST may allows for a more refined and data-driven approach to improving material predictions.

\section{Data Availability}

The test crystal structures are downloaded from the Materials Project database at \url{materialsproject.org} according to the Ids of the test structures as defined in the CSPBenchmark study. The source code of TCSP 2.0 can be found at \url{https://github.com/usccolumbia/TCSP}

\section{Contribution}
Conceptualization, J.H.; methodology,L.W., J.H., R.D., N.F., S.O.; software, L.W.; resources, J.H.; writing--original draft preparation, L.W., J.H., R.D., N.F., S.O.; writing--review and editing, J.H., R.D., N.F., S.O.; visualization, L.W.; supervision, J.H.;  funding acquisition, J.H.

\section*{Acknowledgement}
The research reported in this work was partially supported by the National Science Foundation under grants 2311202 and 2320292. The views, perspectives and content do not necessarily represent the official views of the NSF.

\bibliographystyle{unsrt}  
\bibliography{references}

\end{document}

\maketitle

\section{Dataset preparation}

Our training and test datasets are prepared using the following process. We obtain 151,707 ICSD CIF structure files, and each CIF is annotated with oxidation states. However, there are many entries such as intermetallic materials are assigned 0 oxidation states (OS) to all their atoms. We thus first exclude all structures with 0 oxidation state assignment. We also exclude those materials with fractional oxidation states or those materials with $>$200 atoms. Since ICSD contains many CIF structure files that neglect the hydrogen atoms, we develop an algorithm to add the hydrogen atoms back to the structures along with their OS. Considering that one composition may correspond to multiple structural phases, we select for the formula the oxidation state assignment that appear most frequently among all the polymorphism phases. We also exclude those materials only with a single element. Finally, we obtain 52,147 formulas with valid oxidation states to which we name it as the OS-ICSD dataset. We then split this dataset into the training set, the validation set, and the test set using the 85\%:5\%:15\% ratio.

To examine whether specialized material families can be used to train a more accurate OS prediction model, we first exclude those formulas with fractional atomic numbers and then select only charge-neutral (CN) formulas from the OS-ICSD's training, validation, and test sets, which generate our OS-ICSD-CN dataset. We further select only oxide materials from the OS-ICSD dataset from its training, validation and test sets to compose our OS-ICSD-oxide dataset. Finally, we compose our OS-ICSD-CN-oxide dataset by selecting charge-neural oxides from OS-ICSD's training, validation, and test sets. Our data construction procedure ensures that all four test sets never overlap with all training and validation sets so that the models trained with four different training sets can be tested using any of the four test sets. The details of the final four datasets are shown in Table \ref{tab:datasets}.

\begin{table*}[ht]
\centering
\caption{Statistics of training and test datasets}
\label{tab:datasets}
\begin{tabular}{
>{\columncolor[HTML]{FFFFFF}}c |
>{\columncolor[HTML]{FFFFFF}}c 
>{\columncolor[HTML]{FFFFFF}}c 
>{\columncolor[HTML]{FFFFFF}}c }
\toprule[1.5pt]
{\color[HTML]{000000} \textbf{Dataset}}          & {\color[HTML]{000000} \textbf{Training}} & {\color[HTML]{000000} \textbf{Validation}} & {\color[HTML]{000000} \textbf{Test}} \\ \hline
{\color[HTML]{000000} OS-ICSD}             & {\color[HTML]{000000} 44324}    & {\color[HTML]{000000} 2608}       & {\color[HTML]{000000} 5215} \\
{\color[HTML]{000000} OS-ICSD-CN}       & {\color[HTML]{000000} 31827}    & {\color[HTML]{000000} 1873}       & {\color[HTML]{000000} 3724} \\
{\color[HTML]{000000} OS-ICSD-oxide}    & {\color[HTML]{000000} 30519}    & {\color[HTML]{000000} 1764}       & {\color[HTML]{000000} 3603} \\
{\color[HTML]{000000} OS-ICSD-CN-oxide} & {\color[HTML]{000000} 20601}    & {\color[HTML]{000000} 1208}        & {\color[HTML]{000000} 2420} \\ 
\bottomrule[1.5pt]
\end{tabular}
\end{table*}

\section{OS Distribution in Datasets}

We list the distributions of oxidation states of each element in our four datasets.

\begin{longtable}
{
>{\columncolor[HTML]{FFFFFF}}c |
>{\columncolor[HTML]{FFFFFF}}c |
>{\columncolor[HTML]{FFFFFF}}c |
>{\columncolor[HTML]{FFFFFF}}c |
>{\columncolor[HTML]{FFFFFF}}c }

\caption{Oxidation state distributions in OS-OCSD, OS-ICSD-CN, OS-ICSD-oxide, and OS-ICSD-CN-oxide datasets} \\
\toprule[1.5pt]
{\color[HTML]{3B2322} }            & {\color[HTML]{3B2322} \textbf{OS-ICSD}} & {\color[HTML]{3B2322} \textbf{OS-ICSD-CN}} & {\color[HTML]{3B2322} \textbf{OS-ICSD-oxide}} & \textbf{OS-ICSD-CN-oxide} \\ \hline
{\color[HTML]{3B2322} \textbf{H}}  & {\color[HTML]{3B2322} 1,-1,0}           & {\color[HTML]{3B2322} 1,-1,0}              & {\color[HTML]{3B2322} 1,-1,0}                 & 1,-1,0                    \\ \hline
{\color[HTML]{3B2322} \textbf{He}} & {\color[HTML]{3B2322} 0}                & {\color[HTML]{3B2322} 0}                   & {\color[HTML]{3B2322} 0}                      & 0                         \\ \hline
{\color[HTML]{3B2322} \textbf{Li}} & {\color[HTML]{3B2322} 1,0}              & {\color[HTML]{3B2322} 1,0}                 & {\color[HTML]{3B2322} 1,0}                    & 1,0                       \\ \hline
\textbf{Be}                        & 2                                       & 2                                          & 2                                             & 2                         \\ \hline
\textbf{B}                         & -3,-2,0,2,-1,3,1                        & -3,-2,0,2,-1,3,1                           & -3,0,2,-1,3,1                                 & -3,0,2,-1,3,1             \\ \hline
\bottomrule[1.5pt]
\end{longtable}

\section{BERTOS Network structure and hyper-parameter}
We list the network hyperparameters and the training parameters of our BERTOS network.

\begin{table*}[th!]
\centering
\caption{Hyperparameters of BERTOS network}

\begin{tabular}{
>{\columncolor[HTML]{FFFFFF}}c 
>{\columncolor[HTML]{FFFFFF}}c |
>{\columncolor[HTML]{FFFFFF}}c 
>{\columncolor[HTML]{FFFFFF}}c }
\toprule[1.5pt]
\multicolumn{2}{c|}{\cellcolor[HTML]{FFFFFF}\textbf{Network Parameters}}                                                          & \multicolumn{2}{c}{\cellcolor[HTML]{FFFFFF}\textbf{Training Parameters}}                                                 \\ \hline
\multicolumn{1}{c|}{\cellcolor[HTML]{FFFFFF}{\color[HTML]{000000} Vocabulary Size}}         & {\color[HTML]{000000} 123} & \multicolumn{1}{c|}{\cellcolor[HTML]{FFFFFF}{\color[HTML]{000000} Batch Size}}    & {\color[HTML]{000000} 256}  \\
\multicolumn{1}{c|}{\cellcolor[HTML]{FFFFFF}{\color[HTML]{000000} Hidden Size}}             & {\color[HTML]{000000} 120} & \multicolumn{1}{c|}{\cellcolor[HTML]{FFFFFF}{\color[HTML]{000000} Epochs}}        & {\color[HTML]{000000} 500}  \\
\multicolumn{1}{c|}{\cellcolor[HTML]{FFFFFF}{\color[HTML]{000000} Max Position Embeddings}} & {\color[HTML]{000000} 200} & \multicolumn{1}{c|}{\cellcolor[HTML]{FFFFFF}{\color[HTML]{000000} Learning Rate}} & {\color[HTML]{000000} 1e-3} \\
\multicolumn{1}{c|}{\cellcolor[HTML]{FFFFFF}{\color[HTML]{000000} Attention Heads}}         & {\color[HTML]{000000} 4}   & \multicolumn{1}{c|}{\cellcolor[HTML]{FFFFFF}{\color[HTML]{000000} }}              & {\color[HTML]{000000} }     \\
\multicolumn{1}{c|}{\cellcolor[HTML]{FFFFFF}{\color[HTML]{000000} Hidden Layer}}            & {\color[HTML]{000000} 12}  & \multicolumn{1}{c|}{\cellcolor[HTML]{FFFFFF}{\color[HTML]{000000} }}              & {\color[HTML]{000000} }     \\ 
\bottomrule[1.5pt]
\end{tabular}
\end{table*}

\section{Figures}
We plot confusion matrices for non-metal and metal element OS predictions.

\FloatBarrier

\section{Case studies of predicting complex oxidation states}

\begin{table}[]
\centering
\caption{Case studies of complex OS predictions}
\label{tab:my-table}
\begin{tabular}{c|c c}
\toprule[1.5pt]
                                             \multicolumn{3}{c}{\textbf{$\geq$1 transitionMetal + $\geq$1 non-metal}} \\ \hline
                                             & \multicolumn{1}{c}{Composition}                                & OS                                       \\ \hline
                                             & \multicolumn{1}{c}{Cr8 P16 O52}                                & \textbf{3,5,-2}                          \\ 
                                             & \multicolumn{1}{c}{Si2 Ag8 O8}                                 & \textbf{4,1,-2}                          \\ 
                                             & \multicolumn{1}{c}{Cs8 Mn4 Te8}                                & 1,2,-2                                   \\  
                                             & \multicolumn{1}{c}{Ag20 Sb4 S16}                               & 1,3,-2                                   \\ 
\multirow{-5}{*}{Ternary}                    & \multicolumn{1}{c}{Cr4 Ni2 Se8}                                & \textbf{3,2,-2}                          \\ \hline
                                             & \multicolumn{1}{c}{Rb4 Sm8 Cu4 S16}                            & 1,3,1,-2                                 \\  
                                             & \multicolumn{1}{c}{Y6 Cu2 Si2 S14}                             & 3,1,4,-2                                 \\ 
                                             & \multicolumn{1}{c}{Mn4 P8 H8 O28}                              & \textbf{2,5,1,-2}                        \\  
                                             & \multicolumn{1}{c}{Ta8 H60 N24 Cl28}                           & 5,1,-3,-1                                \\  
\multirow{-5}{*}{Quartenary}                 & \multicolumn{1}{c}{K8 Ti8 P8 O40}                              & 1,4,5,-2                                 \\ \hline
                                             & \multicolumn{1}{c}{Mg4 Hg8 H48 Br24 O24}                       & 2,2,1,-1,-2                              \\  
                                             & \multicolumn{1}{c}{Na12 Ti6 Fe6 P18 O72}                       & 1,4,3,5,-2                               \\ 
                                             & \multicolumn{1}{c}{Cs6 Mn2 Cd6 P24 O72}                        & 1,3,2,5,-2                               \\  
                                             & \multicolumn{1}{c}{In4 Cu2 Ag2 Te4 Se4}                        & 3,1,1,-2,-2                              \\  
\multirow{-5}{*}{\textgreater{}= 5 elements} & \multicolumn{1}{c}{U4 Ag3 P4 H25 O36}                          & \textbf{6,1,5,1,-2}                      \\ 
\bottomrule[1.5pt]
\end{tabular}
\end{table}

\section{Hypothetical material compositions generated by BLMM and filtered by BERTOS}

\begin{lstlisting}
Top 1000 recommended material compositions by BERTOS

composition	formation_energy_preded(eV)
Cs3CrF7	-6.786247253
Cs3CrF6	-6.019856453
Cs3CrF8	-5.872228622


\end{lstlisting}